\documentclass[11pt]{article}

\usepackage[]{acl}

\usepackage{times}
\usepackage{latexsym}
\usepackage{enumitem}
\usepackage[T1]{fontenc}

\usepackage[utf8]{inputenc}

\usepackage{microtype}

\usepackage{inconsolata}

\usepackage{graphicx}

\usepackage{amsmath, amssymb}
\usepackage{mathtools}

\usepackage{booktabs}
\usepackage{multirow}
\usepackage{makecell}
\usepackage{graphicx}
\usepackage[table]{xcolor}

\usepackage{xspace}

\usepackage[most]{tcolorbox}

\newcommand{\ours}{SHIFT\xspace}

\title{SHIFT: Self-reconstruction Harnesses Implicit Fine-grained Thinking for Retrieval}

\author{
Yuxiao Luo\textsuperscript{\rm 1}\thanks{Equal contribution.}\space\space
Da Li\textsuperscript{\rm 2}\footnotemark[1]\space\space
Mingjie Zhang\textsuperscript{\rm 1}\footnotemark[1]\space\space
Zhentao He\textsuperscript{\rm 3}\space\space \\
\textbf{Shikun Zhang}\textsuperscript{\rm 1}\space\space
\textbf{Wei Ye}\textsuperscript{\rm 1}\thanks{Corresponding author.}\space\space\\
\textsuperscript{\rm 1}Peking University, \,\, \textsuperscript{\rm 2}Institute of Computing Technology, Chinese Academy of Sciences\\ \textsuperscript{\rm 3}Beihang University \\
\textsuperscript{}\{luoyuxiao@stu.pku.edu.cn, mjzhang0621@stu.pku.edu.cn, wye@pku.edu.cn\} \\
\textsuperscript{}\{lida21s\}@ict.ac.cn \\
}

\begin{document}
\maketitle
\begin{abstract}
LLM-based retrievers have become a fundamental component of modern information retrieval systems. The paradigm of ``rewrite-then-retrieve'' introduces explicit reasoning before retrieval. In addition, implicit-reasoning retrievers such as GIRCSE and LaSER improve efficiency by replacing explicit reasoning with soft tokens. Although these methods demonstrated competitive performance on reasoning-intensive retrieval benchmarks, they struggle to address the mismatch between the objectives of retrieval and generation. In this work, we propose \textbf{\ours} (\textbf{S}elf-reconstruction \textbf{H}arnesses \textbf{I}mplicit \textbf{F}ine-grained \textbf{T}hinking for Retrieval), a retrieval training framework based on LLMs. Firstly, we transfer LLMs into reasoning-efficient retrievers with residual projection and task-oriented bidirectional attention aggregation in the latent space.
Secondly, we alleviate the mismatch between contrastive learning and implicit reasoning using fine-grained next-token-prediction-based reconstruction. Extensive experiments on reasoning-intensive retrieval benchmarks show that \ours consistently outperforms other widely used retrievers. We also carried out a detailed analysis to illustrate how our method works.
\end{abstract}

\section{Introduction}

Text retrieval has become a core component of modern information retrieval systems, serving a range of applications including web search, open-domain question answering, and retrieval-augmented generation~\citep{dense-passage-retrieval-for-open-domain-question-2020,retrieval-augmented-generation-for-knowledge-int-2020,guu2020realm}. Recently, LLMs demonstrated competitive performance on this task. but existing studies universally treat LLMs as encoders, training them with contrastive objectives to optimize the discriminability of output embeddings~\citep{dense-passage-retrieval-for-open-domain-question-2020,gao2021simcse}. By contrast, the reasoning ability inherited from LLMs remains underexploited in this paradigm.


To leverage the potential of LLMs in retrieval, some work has started to explore how to leverage LLM reasoning to boost retrieval performance, especially for complex retrieval tasks that involve implicit intent and semantic ambiguity to identify relevant content~\citep{bright-a-realistic-and-challenging-benchmark-for-2025}. Building on the ``rewrite-then-retrieve'' paradigm, they first employed LLMs to rewrite or expand queries. Subsequently, the expanded content, together with the original input, was fed into the retriever~\citep{query2doc-query-expansion-with-large-language-mo-2023,chain-of-thought-prompting-elicits-reasoning-in--2022,tongsearch-qr-reinforced-query-reasoning-for-ret-2025}. Although these methods enhance the performance of retrieval tasks, the cascaded retrieval pipeline introduces additional computational overhead and makes it difficult to optimise the retrieval end-to-end. Whether it is possible to leverage the inherent capabilities of LLMs to incorporate rewriting into the process of converting input into embeddings is a question under exploration~\citep{tang2025think_before_rec,jin2026laser}.


Treating the process of expansion or interpretation as a process of reasoning is the primary approach in existing explorations. Based on the form of reasoning, the current work is categorised into explicit and implicit reasoning processes. Each presents its own problems to be resolved.
Explicit reasoning is not an efficient interface. Expressing abstract reasoning with tokens may be verbose and lossy. The reasoning aligns with human cognition, which does not help LLMs to think. 
Implicit reasoning uses a continuous representation in a latent space to perform reasoning, replacing specific tokens~\citep{hao2025coconut, deng2023implicit, shen-etal-2025-codi}, which can reduce reasoning cost, but the reasoning process is difficult to trace. There are also some studies that attempt to use latent space~\citep{li2025coma, chen2026reconstructing}.




However, simply introducing reasoning during encoding does not make it beneficial for retrieval. We identify two mismatches that limit existing methods. 
Firstly, there is a \textbf{representation mismatch}: hidden states produced by causal LLMs are primarily optimized to summarize preceding context for next-token generation, whereas dense retrieval requires representations that support query-document matching and discriminate relevant documents from hard negatives. Directly using or aligning these states may preserve generation-oriented redundancy~\citep{ethayarajh2019contextual, gao2019representation, wang2020spectrum} rather than retrieval-relevant reasoning, causing different latent steps to collapse into similar static embeddings. Secondly, there is a \textbf{supervision mismatch}: standard contrastive learning supervises only the final query-document similarity, providing little guidance on what each intermediate reasoning state should encode~\citep{liu2026crem}. As a result, reasoning tokens may be introduced architecturally but remain weakly shaped semantically.

To resolve the mismatch mentioned above, we introduce \textbf{\ours} (\textbf{S}elf-reconstruction \textbf{H}arnesses \textbf{I}mplicit \textbf{F}ine-grained \textbf{T}hinking for Retrieval) in this work. On the representation, \ours transforms latent reasoning states into retrieval-oriented representations through residual projection, which filters generation-oriented contextual noise, and bidirectional attention aggregation, which dynamically combines multiple latent reasoning steps without assuming that later steps are always more useful. As for training, \ours~introduces fine-grained self-reconstruction based on next-token prediction. Instead of directly aligning implicit reasoning with explicit reasoning, which may inherit their redundancy, we expand step-level latent states into token-level representations and train them to reconstruct explicit reasoning trajectories.

We conducted experiments on several representative reasoning-intensive retrieval benchmarks to demonstrate the effectiveness of \ours. The experimental results demonstrate that \ours~ is effective, achieving optimal performance when using the same LLM as the backbone. At the same time, it requires only a three-step reasoning process, significantly reducing the computational overhead introduced during the encoding.
Further analysis of latent representations in the reasoning process reveals that each latent representation can improve retrieval performance monotonically, since different reasoning steps may capture different retrieval intents. These representations require dynamic integration to achieve further performance improvements.

In summary, our contributions are as follows:
\begin{itemize}[leftmargin=*]
    \item We propose \ours, a framework that transfers LLMs into efficient reasoning-intensive retrievers. Extensive experiments demonstrate that \ours outperforms baselines.
    \item We systematically study how implicit representations can be used for retrieval, comparing multiple transformations with both empirical results and theoretical analysis.
    \item We analyze the effectiveness of supervision and discuss why our supervision better shapes meaningful latent reasoning space.
\end{itemize}

\section{Related Works}

\subsection{LLM-based Retrieval with Explicit Rewriting and Reasoning}

Dense retrieval has evolved from discriminative encoders such as BERT~\citep{bert-pre-training-of-deep-bidirectional-transfor-2019} to retrievers built on top of generative LLM backbones~\citep{improving-text-embeddings-with-large-language-mo-2024, your-dense-retriever-is-secretly-an-expeditious--2025b}. While these models improve general semantic matching, they are still typically trained with contrastive objectives over final representations, which can be insufficient for reasoning-intensive or ambiguity-heavy queries. A common strategy is to introduce \emph{explicit} intermediate text before retrieval, either by rewriting or expanding the query~\citep{precise-zero-shot-dense-retrieval-without-releva-2023, query2doc-query-expansion-with-large-language-mo-2023} or by generating reasoning-oriented rationales and auxiliary textual signals~\citep{c-pack-packed-resources-for-general-chinese-embe-2024, tongsearch-qr-reinforced-query-reasoning-for-ret-2025, think-then-embed-generative-context-improves-mul-2025, answer-is-all-you-need-instruction-following-tex-2024, grace-generative-representation-learning-via-con-2025, expandr-teaching-dense-retrievers-beyond-queries-2025}. Although effective, these methods increase huge latency and depend on a strong reasoner optimized towards the retrieval task.




\subsection{Implicit Reasoning in Retrieval}
Recent work on \emph{latent reasoning} moves intermediate computation from token space to continuous hidden states \citep{hao2025coconut, liu2024hidden, shen-etal-2025-codi, tan2025colar, zeng2026adaptive}. Unlike explicit text spaces, introducing effective supervision into implicit spaces is a challenging problem. 
Implicit reasoning in retrieval remains much less explored than its counterpart in general LLM reasoning. GIRCSE \citep{let-llms-speak-embedding-languages-generative-te-2025} shows that a retriever can iteratively refine its representation through autoregressive generation of soft embedding tokens. LaSER \citep{jin2026laser} uses explicit reasoning paths as privileged signals and distills them into the latent space of dense retrievers through a dual-view framework, which aligns not only final outputs but also intermediate reasoning steps. 


\section{Methodology}

\begin{figure*}[t]
    \centering
    \includegraphics[width=\textwidth]{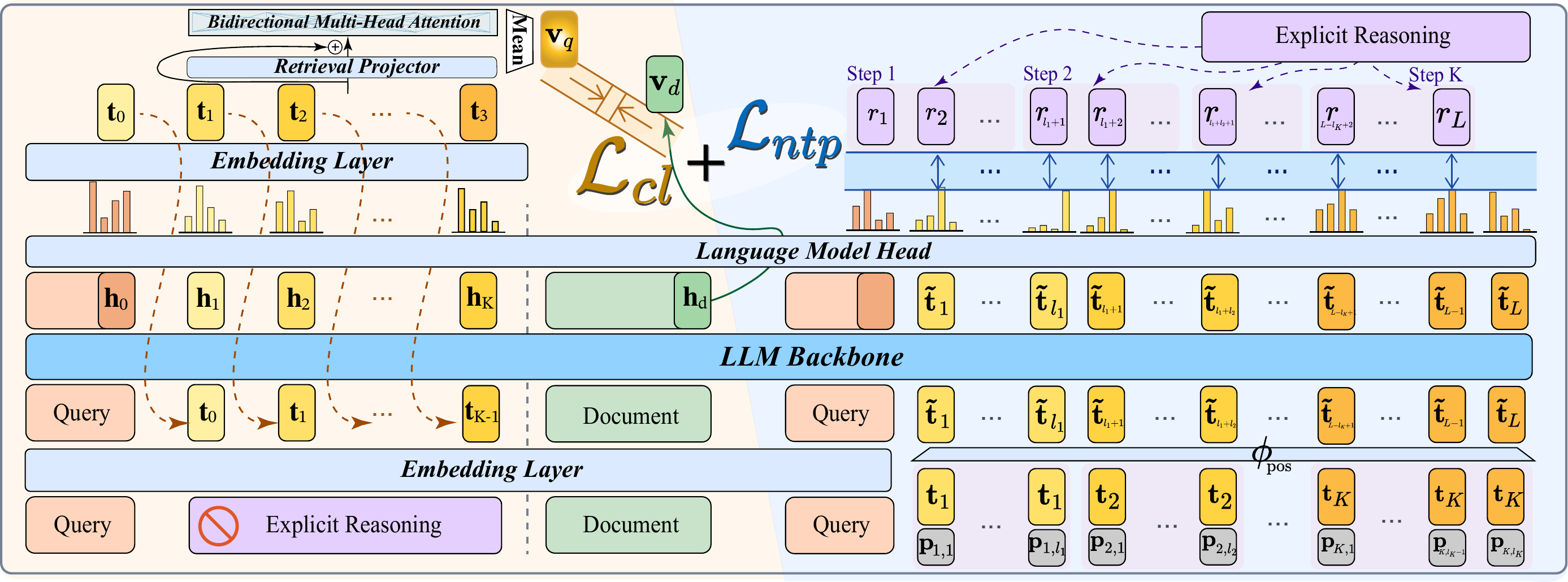}
    \caption{Main architecture of \ours. During training, we supervise the latent representations by both \textbf{Contrastive Learning (CL)} Loss (shown on the left) and \textbf{Next-Token-Prediction (NTP)} Loss (shown on the right). \textbf{CL} loss shapes the retrieval space by pushing and pulling retrieval targets, while \textbf{NTP} loss shapes the latent reasoning space by autoregressive self-reconstruction. }
    \label{fig:main}
\end{figure*}

\subsection{Task Formulation}
Given a user query $q$ and a candidate document $\mathcal{D}=\{d_1,d_2,\ldots,d_N\}$, retrieval aims to learn an encoder $f(\cdot)$ that projects textual inputs into an $m$-dimensional embedding space $\mathbb{R}^m$. The relevance between a query $q$ and a document $d$ is commonly measured by the cosine similarity between their corresponding embeddings $\mathbf{v}_q$ and $\mathbf{v}_d$. In conventional LLM-based retrievers, the input embedding is obtained as the hidden state of the last layer of the \texttt{<EOS>} token, which is appended to the input sequence: 
$
  \mathbf{v}_{\mathrm{input}} = f(\mathrm{Input}; \texttt{<EOS>})
$. To solve reasoning-intensive queries, existing ``rewrite-then-retrieve'' methods \cite{reasonembed-enhanced-text-embeddings-for-reasoni-2025b} introduce an external LLM reasoner $\mathcal{R}$ to first produce an explicit reasoning trace $r_q$, which is subsequently concatenated with the original query and fed into the retriever:
$
  \mathbf{v}_q^* = f([q; R_q; \texttt{<EOS>}]), \  
  \text{where } R_q \sim \mathcal{M_{R}}(q).
$

\subsection{Overview}

As Figure~\ref{fig:main}~shows, we build \ours~upon an LLM with causal attention. First, \ours~autoregressively produces a sequence of $K$ continuous latent thinking tokens. Then these latent states will be transformed to a new latent space for retrieval via a retrieval head. The final query representation $v_q$ is then computed from these transformed latent states using gated attention pooling.

\subsection{Soft Thinking in Latent Continuous Space}

Motivated by recent progress in continuous reasoning, we introduce a mechanism where the encoder $f_{\theta}(\cdot)$, initialized from a causal LLM, encodes the sequence inputs into a single embedding via the last hidden states of the last token. As illustrated in Figure~\ref{fig:main}, given an input query $q$, we directly get the input embeddings $X=\mathbf{E}(q)$ where $\mathbf{E} \in \mathbb{R}^{|V| \times m}$ denotes the embedding matrix and $|V|$ is the vocabulary size. 

Unlike standard decoding, which selects discrete vocabulary indices, latent tokens are represented as continuous embeddings to preserve semantic richness while maintaining differentiability. The generation process is autoregressive: at the $j$-th reasoning step ($1 \leq j \leq K$), let $\mathbf{h}_{j-1} \in \mathbb{R}^m$ denote the last hidden state produced by the backbone where $\mathbf{h}_0=f_{\theta}(\mathbf{E}(q))$, we project $\mathbf{h}_{j-1}$ to the vocabulary space using the language modeling head $\mathbf{W}_{\mathrm{lm}} \in \mathbb{R}^{m \times |V|}$, yielding
\begin{align}
  \mathbf{logit}_j = \mathbf{W}_{\mathrm{lm}} \mathbf{h}_{j-1}.
\end{align}
Rather than applying a hard decoding operation such as $\arg\max$, we compute a vocabulary distribution:
\begin{align}
  \mathbf{p}_j = \mathrm{softmax}(\mathbf{logit}_j).
\end{align}
The soft latent token $t_j$ is defined as the expectation of the embedding vectors under this distribution:
\begin{align}
  \mathbf{t}_j = \mathbf{p}_j^T \mathbf{E} = \bigl(\mathrm{softmax}(\mathbf{W_{\mathrm{lm}} h}_{j-1})\bigr)^T \mathbf{E}.
\end{align}
The resulting soft token $t_j$ is appended to the input sequence for the subsequent reasoning step. Accordingly, at step $j$, the model takes as input the original query embeddings together with all previously generated latent tokens:
\begin{align}
  \mathbf{h}_j = f_\theta([\mathbf{E}(q); \mathbf{t}_1, \ldots, \mathbf{t}_j]).
\end{align}
Through causal attention, each latent token $t_j$ is able to attend to both the query context and earlier reasoning steps, thereby resembling the sequential dependency pattern of explicit Chain-of-Thought reasoning.

\subsection{Residual Projected Latent Retrieval Space}


After $K$ steps, we obtain a sequence of latent reasoning states
$
    \mathbf{H}_{latent} = [\mathbf{h}_1, \ldots, \mathbf{h}_K], \quad \mathbf{h}_t \in \mathbb{R}^m.
$
We first apply a lightweight two-layer residual projection $\phi$ to obtain retrieval-oriented step latent representations while filtering the common contextual information prepared for generation, which will be discussed in section \ref{sec:anisotropies}.
\begin{align}
    \mathbf{\tilde{h}}_i = \mathbf{h}_i + \phi_{\mathrm{ret}}(\mathbf{h}_i),
\end{align}
where $\phi_{\mathrm{ret}}(\cdot)$ is a learnable projection module.

\subsection{Bi-Attention Pooling}

The required reasoning depth varies across queries. Although we fix the number of latent reasoning steps to $K$ for simpler supervision, different tasks may still rely on different subsets of these latent representations.
Therefore, we introduce an attention-based pooling module to adaptively aggregate the implicit reasoning states (The choice of aggregating strategies is discussed in \ref{sec:agg}). Obtained the latent reasoning representations
  $\tilde{\mathbf{H}}=[\tilde{\mathbf{h}}_1,\ldots,\tilde{\mathbf{h}}_K]\in\mathbb{R}^{K\times m}$. We use a self-attention mechanism to assign different weights for different reasoning steps:
\begin{equation}
\begin{aligned}
  \mathbf{Q}_i = \tilde{\mathbf{H}}&\mathbf{W}^{Q}_i, 
  \mathbf{K}_i = \tilde{\mathbf{H}}\mathbf{W}^{K}_i, 
  \mathbf{V}_i = \tilde{\mathbf{H}}\mathbf{W}^{V}_i, \\
  \mathbf{o}_i &=
  \mathrm{softmax}
  \left(
  \frac{\mathbf{Q}_i\mathbf{K}_i^{\top}}{\sqrt{m/n_{\mathrm{h}}}}
  \right)
  \mathbf{V}_i, \\
  \mathbf{z} &=
  \mathrm{concat}(\mathbf{o}_1,\ldots,\mathbf{o}_{n_{\mathrm{h}}})\mathbf{W}^{O},
\end{aligned}
\end{equation}
where $n_{\mathrm{h}}$ is the number of attention heads, and $\mathbf{z}\in\mathbb{R}^{K\times m}$ is the attention-interacted reasoning representations used for retrieval. We finally obtain the pooled query embedding by mean pooling:
\begin{equation}
\begin{aligned}
    \mathbf{v}_q = \frac{1}{K}\sum_{i=1}^{K}\mathbf{z_i}.
\end{aligned}
\end{equation}

For documents, we extract their embeddings just from the last hidden layer of \texttt{<EOS>} without reasoning, as there is no complex instruction for encoding a static document. We will show in Section~\ref{sec:ab_doc} that there are no benefits to doing this and actually increases computational overhead. So the document embedding is 
\begin{equation}
\begin{aligned}
    \mathbf{v}_d = f_{\theta}([\mathbf{E}(d); \texttt{<EOS>}]).
\end{aligned}
\end{equation}
And we will calculate the InfoNCE loss to conduct contrastive learning:

\begin{align}
    \mathcal{L}_{cl} = -\log\frac{\exp\big({\text{sim}(\mathbf{v}_q, \mathbf{v}_{d^{+}})/\tau\big)}}{\sum_{d\in\mathcal{B}}\exp\big({\text{sim}(\mathbf{v}_q, \mathbf{v}_{d})/\tau\big)}},
\end{align}
where $\tau$ is the temperature hyperparameter and $\mathcal{B}$ is a training batch.

\subsection{Fine-grained Self-Reconstruction}


Specifically, as shown in Figure \ref{fig:main}, we expand the step-level latent representations to align the token-level explicit reasoning steps. Formally, we denote the explicit reasoning tokens with $K$ steps as $R_q = \{\{r_1, \cdots, r_{l_1}\}, \cdots, \{r_{L-l_{K}+1}, \cdots, r_L\}\}$, then we can obtain the expanded latent representations via
\begin{align}
  \mathbf{T}_{\mathrm{exp}}
  = \{\underbrace{\mathbf{t}_1, \cdots, \mathbf{t}_1}_{\text{$l_1$ times}}, \cdots, \underbrace{\mathbf{t}_K, \cdots, \mathbf{t}_K}_{l_{K}\text{ times}}\}.
\end{align}

To introduce more meaningful positional information, we combine step-level and token-level information into a 2D reason-wise position embedding. Formally, given the $k$-th step representation $t_{k}, k=1,2,\cdots, K$, we define the normalized step position $\alpha_k$ and the normalized intra-step token position for the $i$-th repetition as $\beta_i$ to compose a 2D positional feature for the expanded token, which is denoted as 
\begin{align}
    \alpha_k = \dfrac{k - 1}{K}, \ \beta_i = \dfrac{i - 1}{l_{i}}, \ 
    \mathbf{p}_{k,i} =
      \begin{bmatrix}
          \alpha_k\\
          \beta_{i}
      \end{bmatrix}.
\end{align}

We concatenate the repeated step representation with the positional feature:
$
    \mathbf{z}_{k, i} =
    \left[
        \mathbf{t}_{k};
        \mathbf{p}_{k,i}
    \right]\in \mathbb{R}^{d+2}
$. Then we apply a small MLP $\phi_{\text{pos}}$, which produces a position-dependent residual. We denote the expanded inputs as $\mathbf{T}_{\mathrm{exp}}\coloneqq \{\tilde{\mathbf{t}}_1, \cdots, \tilde{\mathbf{t}}_L\}$. Finally, the expanded token embedding is given by
\begin{align}
    \mathbf{\tilde{t}}_{\sum_{j=1}^{k-1}l_j + i} = \mathbf{t}_k + \phi_{\text{pos}}\big(\left[
        \mathbf{t}_{k};
        \mathbf{p}_{k,i}
    \right]\big) \in \mathbb{R}^d.
\end{align}

In practice, as we fix the number of latent reasoning steps to $K$ for simplicity, we accordingly uniformly split the explicit reasoning tokens into $K$ segments. 

Then we ask the model to predict the next explicit token given the latent representations. Such a prediction task pushes the model to expand the step-level latent reasoning state to a fine-grained token-level semantics space, shaping the latent space of reasoning states under the supervision of a specific explicit reasoning trajectory using the intrinsic capability of autoregressive generation. We calculate the next-token-prediction loss to the model to map the step-level latent representations to fine-grained token-level reasoning tokens.

\begin{align}
      \mathcal{L}_{ntp} = -\sum_{i=1}^{L} \log\mathrm{P}(r_i\ |\ I,q,\tilde{t}_{<i};\  \theta)
\end{align}

Finally, the model parameters $\theta$ are updated end-to-end to minimize the combined objective
\begin{align}
    \mathcal{L} = \mathcal{L}_{cl} + \mathcal{L}_{ntp}
\end{align}

\section{Experiments}
\subsection{Datasets and Evaluation}

We train \ours on ReasonEmbed~\citep{reasonembed-enhanced-text-embeddings-for-reasoni-2025b},  which is a reasoning-intensive dataset that contains 81K training examples from 12 domains. Each query is paired with a reasoning path generated by GPT-4o-mini \citep{gpt-4o-system-card-2024}. For evaluation, we keep the same settings as LaSER. We use three reasoning-intensive benchmarks: Bright~\citep{bright-a-realistic-and-challenging-benchmark-for-2025} for in-domain evaluation, and two out-of-domain benchmarks: FollowIR~\citep{followir-evaluating-and-teaching-information-ret-2025} and BrowseComp-Plus~\citep{browsecomp-plus-a-more-fair-and-transparent-eval-2025d}. The statistics of datasets are shown in Table~\ref{tab:dataset_stats}.
\begin{table}[htbp!]
\caption{Statistics of the datasets used in our experiments. “Q” and “C” denote query and corpus, respectively. “Len.” denotes average token count. “R. Len.” denotes average reasoning length.}
\label{tab:dataset_stats}
\resizebox{\columnwidth}{!}{%
\begin{tabular}{l l r r r r}
\toprule
 & Dataset & \# Q & \# C & Q Len. & Reas. Len. \\
\midrule
\textit{Train} & ReasonEmb & 81,659 & -- & 222.1 & 984.8 \\
\midrule
\multirow{3}{*}{\textit{Test}}
& Bright (ID)      & 1,384 & 1,145,164 & 240.8 & 2,412.8 \\
& FollowIR (OOD)   & 104   & 98,312    & 76.6  & -- \\
& BC-Plus (OOD)    & 830   & 100,195   & 123.2 & -- \\
\bottomrule
\end{tabular}%
}
\end{table}

Following the evaluation in LaSER, we report nDCG@10 for Bright, and Recall@5, Recall@100, and Recall@1000 for BrowseComp-Plus. For FollowIR, each subset is evaluated with both a standard retrieval metric and p-MRR, a pairwise metric designed to measure instruction-following ability. Specifically, in addition to p-MRR, we report MAP@5 for the Robust'04 and Core'17 subsets, and nDCG@5 for the News'21 subset.

\subsection{Baselines}

We compare \ours with four groups of retrievers reported by LaSER. Unless otherwise specified, trainable baselines use the same backbone and training data as our model.

\textbf{Dense retrievers.} We evaluate both BERT-based and LLM-based retrievers, including BGE-M3 \citep{chen2024bge-m3}, E5-Large-Instruct \citep{text-embeddings-by-weakly-supervised-contrastive-2022} and Qwen3-Embedding \citep{qwen3-embedding-advancing-text-embedding-and-rer-2025a}. We also train a fair contrastive baseline on our composite training set, which isolates the gain from our architecture.

\textbf{Rewrite-then-retrieve pipelines.} This group explicitly rewrites the input query before retrieval. We use BGE-Reasoner-Rewriter-7B from ReasonEmbed \citep{reasonembed-enhanced-text-embeddings-for-reasoni-2025b} to generate rewritten queries, which are then encoded by the fair baseline retriever at inference time.

\textbf{Explicit-reasoning retrievers.} These methods integrate generation and retrieval in a single model by first producing explicit reasoning traces or auxiliary textual outputs before forming the final representation. We compare with Search-R3~\citep{search-r3-unifying-reasoning-and-embedding-gener-2025} and GRACE~\citep{grace-generative-representation-learning-via-con-2025}. 

\textbf{Implicit-reasoning retrievers.} Finally, we compare with GIRCSE~\citep{let-llms-speak-embedding-languages-generative-te-2025} and LaSER~\citep{jin2026laser}, which use latent tokens to perform intermediate reasoning before producing retrieval embeddings.

\begin{table*}[t]
\centering
\small
\setlength{\tabcolsep}{3.7pt}
\renewcommand{\arraystretch}{1.12}
\caption{Main retrieval performance on the Bright benchmark. We report nDCG@10 for all subsets. The best results among models with the same backbone are formatted in bold. $\dagger$ indicates the pipeline method that uses an external LLM to rewrite queries during inference.}
\resizebox{\textwidth}{!}{%
\begin{tabular}{l  r c c c c c c c c c c c c  c}
\toprule
\multirow{2}{*}{\textbf{Models}} & \multirow{2}{*}{\textbf{Size}} &
\multicolumn{7}{c}{\textbf{StackExchange}} &
\multicolumn{2}{c}{\textbf{Coding}} &
\multicolumn{3}{c}{\textbf{Theorem-based}} &
\multirow{2}{*}{\textbf{Avg.}} \\
\cmidrule(lr){3-9} \cmidrule(lr){10-11} \cmidrule(lr){12-14}
& &
\textbf{Bio.} & \textbf{Earth.} & \textbf{Econ.} & \textbf{Psy.} & \textbf{Rob.} & \textbf{Stack.} & \textbf{Sus.} &
\textbf{Leet.} & \textbf{Pony} &
\textbf{AoPS} & \textbf{TheoQ.} & \textbf{TheoT.} &
\\
\midrule
\multicolumn{15}{l}{\textit{\textbf{Standard Dense Retrievers}}} \\
BGE-M3 & 0.6B & 7.2 & 13.3 & 12.5 & 12.8 & 12.6 & 10.0 & 10.1 & 15.6 & 30.7 & 1.5 & 5.6 & 5.0 & 11.4 \\
Multilingual-E5-Large-Instruct & 0.6B & 15.0 & 24.6 & 14.0 & 14.6 & 16.1 & 10.3 & 13.8 & 15.1 & 2.1 & 3.0 & 11.6 & 6.9 & 12.3 \\
Qwen3-Embedding-0.6B & 0.6B & 12.7 & 26.3 & 17.9 & 16.5 & 12.5 & 12.4 & 12.2 & \textbf{14.3} & 0.7 & 3.1 & 17.2 & \textbf{26.5} & 14.4 \\
Qwen3-Embedding-8B & 8B & 14.7 & 17.9 & 15.5 & 19.9 & 9.1 & 12.9 & 16.5 & \textbf{17.4} & 0.8 & 2.5 & 16.8 & 24.5 & 14.0 \\
\midrule
\multicolumn{15}{l}{\textit{\textbf{Basic Contrastive Learning}}} \\
Fair Baseline (Qwen3-0.6B) & 0.6B & 28.8 & 31.6 & 25.2 & 28.1 & 15.1 & 22.3 & 24.2 & 8.8 & 3.5 & 2.1 & 14.1 & 15.4 & 18.3 \\
Fair Baseline (Qwen3-8B) & 8B & 49.7 & 51.2 & 26.9 & 37.4 & 23.4 & 28.0 & 34.1 & 3.7 & 3.2 & 2.8 & 16.8 & 31.8 & 25.7 \\
Fair Baseline (LLaMA3.1-8B) & 8B & 57.7 & 40.9 & 24.6 & 29.8 & 20.1 & 29.2 & 25.8 & 4.3 & 5.7 & 1.6 & 12.0 & 17.7 & 22.5 \\
\midrule
\multicolumn{15}{l}{\textit{\textbf{Explicit Reasoning}}} \\
Rewrite-then-Retrieve (Qwen3-0.6B) $\dagger$ & 0.6B & \textbf{57.8} & \textbf{51.6} & 14.1 & \textbf{39.4} & 15.0 & 19.8 & 23.7 & 0.6 & \textbf{14.6} & 1.2 & 14.9 & 15.5 & 22.4 \\
Rewrite-then-Retrieve (Qwen3-8B) $\dagger$ & 8B & 53.1 & 54.3 & \textbf{32.1} & 34.8 & 20.5 & 31.1 & 32.2 & 3.2 & \textbf{15.2} & 4.1 & 17.4 & \textbf{38.8} & 28.1 \\
Rewrite-then-Retrieve (LLaMA3.1-8B) $\dagger$ & 8B & \textbf{63.1} & \textbf{54.5} & 27.7 & 40.6 & 17.1 & 28.1 & 28.2 & 1.6 & \textbf{11.4} & 3.8 & 16.1 & \textbf{30.3} & \textbf{26.9} \\
Search-R3 (Qwen2.5-1.5B) & 1.5B & 13.8 & 8.3 & 4.2 & 3.5 & 4.5 & 12.4 & 4.6 & 16.7 & 3.0 & 0.6 & 8.5 & 11.7 & 7.7 \\
\midrule
\multicolumn{15}{l}{\textit{\textbf{Latent Reasoning}}} \\
GIRCSE (Qwen3-0.6B) & 0.6B & 29.0 & 32.8 & 24.7 & 30.6 & 13.6 & 24.0 & 26.6 & 11.1 & 1.1 & 1.3 & 13.5 & 20.6 & 19.1 \\
GIRCSE (Qwen3-8B) & 8B & \textbf{59.0} & \textbf{56.5} & 27.2 & 40.3 & 19.0 & 28.5 & 31.4 & 3.2 & 3.6 & 1.7 & 14.0 & 27.2 & 26.0 \\
GIRCSE (LLaMA3.1-8B) & 8B & 50.6 & 43.4 & 28.3 & 35.7 & 14.3 & 26.9 & 26.4 & \textbf{6.0} & 5.2 & 0.5 & 11.6 & 15.1 & 22.0 \\
LaSER (Qwen3-0.6B) & 0.6B & 50.0 & 45.9 & 25.7 & 32.4 & 18.4 & 27.1 & 26.5 & 9.1 & 2.7 & 1.2 & 16.0 & 22.4 & 23.1 \\
LaSER (Qwen3-8B) & 8B & 58.0 & 51.8 & 29.0 & \textbf{44.1} & 26.0 & 32.9 & \textbf{34.5} & 9.2 & 11.7 & 1.6 & 18.7 & 34.0 & 29.3 \\
LaSER (LLaMA3.1-8B) & 8B & 58.4 & 48.1 & 28.0 & \textbf{40.9} & 17.0 & 29.9 & 28.3 & 1.7 & 5.9 & 1.5 & 14.6 & 19.2 & 24.4 \\ 
\rowcolor{blue!10}
\multicolumn{15}{l}{\textit{\textbf{Ours}}} \\
\rowcolor{blue!10}
\textbf{\ours} (Qwen3-0.6B) & 0.6B & 49.8 & 48.7 & \textbf{29.0} & 29.0 & \textbf{20.5} & \textbf{28.5} & \textbf{27.1} & 12.3 & 5.3 & \textbf{4.4} & \textbf{21.8} & 23.6 & \textbf{25.0} \\
\rowcolor{blue!10}
\textbf{\ours} (Qwen3-8B) & 8B & 54.8 & 49.4 & 32.0 & 42.4 & \textbf{26.7} & \textbf{35.2} & 34.3 & 14.2 & 14.1 & \textbf{7.1} & \textbf{26.6} & 35.6 & \textbf{31.0} \\
\rowcolor{blue!10}
\textbf{\ours} (LLaMA3.1-8B) & 8B & 54.4 & 46.4 & \textbf{31.7} & 40.3 & \textbf{24.0} & \textbf{33.0} & \textbf{30.6} & 5.6 & 7.1 & \textbf{4.5} & \textbf{16.8} & 21.2 & 26.3 \\
\bottomrule
\end{tabular}%
}
\label{tab:main}
\end{table*}

\subsection{Implementation Details}

All models are trained for one epoch with LoRA ($r=64$, $\alpha=32$) on 4 A100 GPUs, using a global batch size of 64. We optimize models with AdamW, using a learning rate of $1\times10^{-4}$ and a warmup ratio of 0.1. 
For contrastive training, each query is paired with one hard negative and uses in-batch negatives. The maximum sequence length for both queries and documents is set to 1024 during training and increased to 8192 during inference to support longer contexts. 
The temperature parameter of the contrastive loss, $\tau$, is set to 0.02. The number of latent thinking steps is fixed to $K=3$ for both training and inference. We adopted the same training setup as LaSER for fair comparison. 
We also conducted experiments on a range of LLMs such as Qwen3 series (0.6B, 4B, and 8B) and LLaMA-3.1-8B to demonstrate the generalisability of our method. 

\section{Overall Performance}



In this section, we provided a comprehensive comparison between \ours and existing retrievers. We show the in-domain evaluation results in Table~\ref{tab:main} and the out-of-domain evaluation in Table \ref{tab:main2}, which is displayed in Appendix~\ref{sec:app_results} due to space constraints. And our key observations are summarized as follows.

First, \ours achieves substantial improvements over the fair baseline trained with standard contrastive learning. As shown in Table \ref{tab:main}, \ours based on Qwen3-8B not only surpasses the original Qwen3-Embedding-8B model, but also outperforms the fine-tuned fair baseline. This clear performance gap indicates that standard contrastive training with a conventional retrieval architecture is insufficient for complex reasoning tasks. 
Second, \ours outperforms existing implicit reasoning methods. We further compare \ours with state-of-the-art implicit reasoning retrievers, particularly LaSER and GIRCSE. The result highlights the effectiveness of our approach.

\section{Further Anaysis}

In this section, we dive into a deep analysis of latent representations. We first conduct a comprehensive ablation study with Qwen3-0.6B to examine the contribution of our components separately. We further provide empirical and analytical justification for the design of \ours, demonstrating how our approaches address the above mismatches.

\subsection{Ablation Study}

As shown in Table \ref{tab:ablation}, \textbf{w/o Reconstruction} removes the self-prediction reconstruction loss, which leads to a performance degradation, demonstrating the effectiveness of the supervision introduced by reconstruction. \textbf{w/o Residual Projection} removes $\phi_{\mathrm{ret}}$, illustrating the importance of distinguishing between retrieval space and reasoning space. \textbf{w/o 2D Position} removes 2D position embeddings, leading to a suboptimal performance.

\begin{table}[h]
\centering

\renewcommand{\arraystretch}{1.14}
\caption{Ablation study of \textbf{\ours} (Qwen3-0.6B) on Bright, FollowIR, and BrowseComp-Plus.}
\resizebox{\columnwidth}{!}{%
\begin{tabular}{l c c c c c}
\toprule
\multirow{2}{*}{\textbf{Method}} &
\multicolumn{1}{c}{\textbf{Bright}} &
\multicolumn{2}{c}{\textbf{FollowIR}} &
\multicolumn{2}{c}{\textbf{BrowseComp-Plus}} \\
\cmidrule(lr){2-2} \cmidrule(lr){3-4} \cmidrule(lr){5-6}
& \textbf{nDCG@10} & \textbf{Score} & \textbf{p-MRR} & \textbf{R@5} & \textbf{R@1000} \\
\midrule
\rowcolor{blue!10}
\textbf{\ours}(Qwen3-0.6B) & \textbf{25.0} & 11.3 & 1.6 & 6.9 & 56.1 \\
\midrule
w/o Reconstruction & 22.1 & 9.8 & -6.7 & 5.5 & 48.6 \\
w/o Residual Projection & 22.8 & 10.1 & -5.4 & 6.0 & 52.8 \\
w/o 2D Position & 24.4 & 11.3 & 1.6 & 6.8 & 56.1 \\
\midrule
Retrieval with <EOS> & 20.3 & 10.6 & 0.0 & 4.1 & 47.2 \\
\midrule
Mean Pooling & 24.3 & 11.1 & 4.5 & 6.8 & 52.9 \\
Last Pooling & 20.9 & 9.7 & 0.8 & 4.7 & 51.0 \\
Casual MHA Pooling & 21.8 & 9.6 & 0.8 & 5.1 & 48.9 \\
GRU Pooling & 15.6 & 5.8 & -5.5 & 3.3 & 46.6 \\
\midrule
BERT Reconstruction & 21.7 & 9.8 & -5.1 & 4.2 & 47.1 \\
Prefix Reconstruction & 23.4 & 11.6 & 0.2 & 6.4 & 52.8 \\
\midrule
w Document Reasoning & 24.9 & \textbf{11.7} & \textbf{2.0} & 6.8 & \textbf{56.2} \\
\midrule
Basic Contrastive Learning. & 18.3 & 7.4 & -0.2 & 3.5 & 46.9 \\
\bottomrule
\end{tabular}%
}
\label{tab:ablation}
\end{table}

\subsection{Latent Space Are Promising for Retrieval}

To study whether latent representations are effective for retrieval, we design a baseline to follow the conventional ``rewrite-then-retrieve'' paradigm where the model generates an explicit reasoning trace or rewritten query followed by the final \texttt{<EOS>} instead of directly transferring latent representations for retrieval. Our experiments, shown in Table \ref{tab:ablation}, illustrate that supervising only the final terminal token wastes the rich information contained in the latent representations. Aggregating latent representations into a single representation \texttt{<EOS>} factually works as a relatively static pooling method. 

\subsection{Less Alignment, Less Redundancy}
\label{sec:anisotropies}



Previous latent reasoning retrievers suffer from a fundamental supervision bottleneck. Although multiple latent reasoning representations are introduced, they are usually supervised only through the final retrieval objective, often after mean pooling. Such supervision does not guide the pushing of different latent states to encode distinct reasoning steps. As a result, the model often minimizes the retrieval loss by making all latent representations converge to a similar vector. 

\begin{figure}[h]
    \centering
    \includegraphics[width=1\columnwidth]{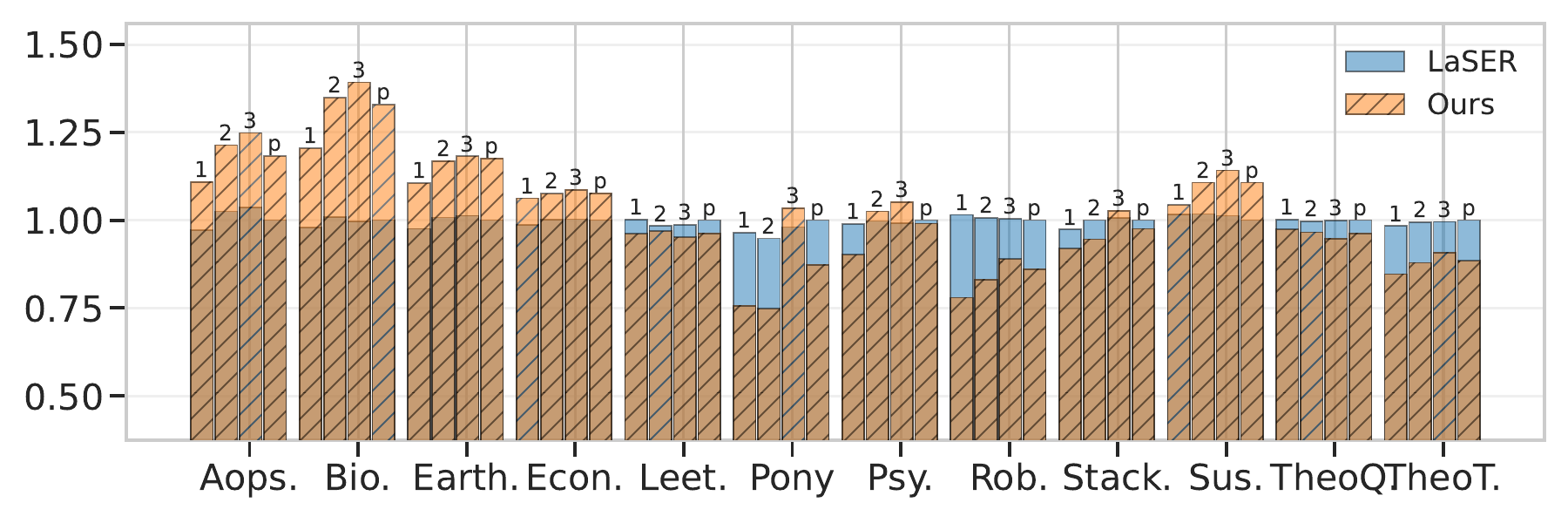}
    \caption{Relevant nDCG@10 ratio of \ours between LaSER. The numbers  \textbf{1, 2, 3} above the bars are the token index, and \textbf{p} means pooled.}
    \label{fig:step_perf}
\end{figure}

We observe this degeneration clearly in LaSER. When evaluating retrieval performance using each of its three latent representations independently, the results are nearly identical, as shown in Figure \ref{fig:step_perf}. 

\begin{figure}[htbp]
    \centering
    \includegraphics[width=1\columnwidth]{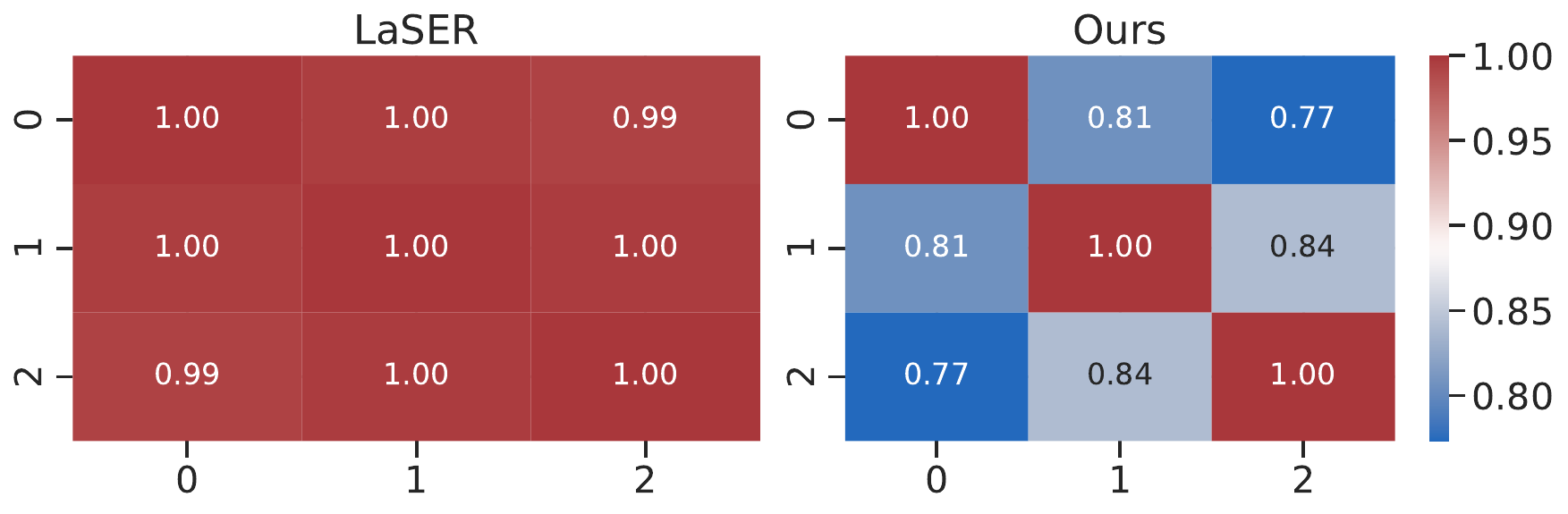}
    \caption{Similarity between latent representations.}
    \label{fig:laser_sim}
\end{figure}

We then discovered that the performance similarity is caused by the indistinguishability of latent states: their pairwise cosine similarities are close to $1$, as shown in Figure \ref{fig:laser_sim}.

\begin{figure}[htbp]
    \centering
    \includegraphics[width=1\columnwidth]{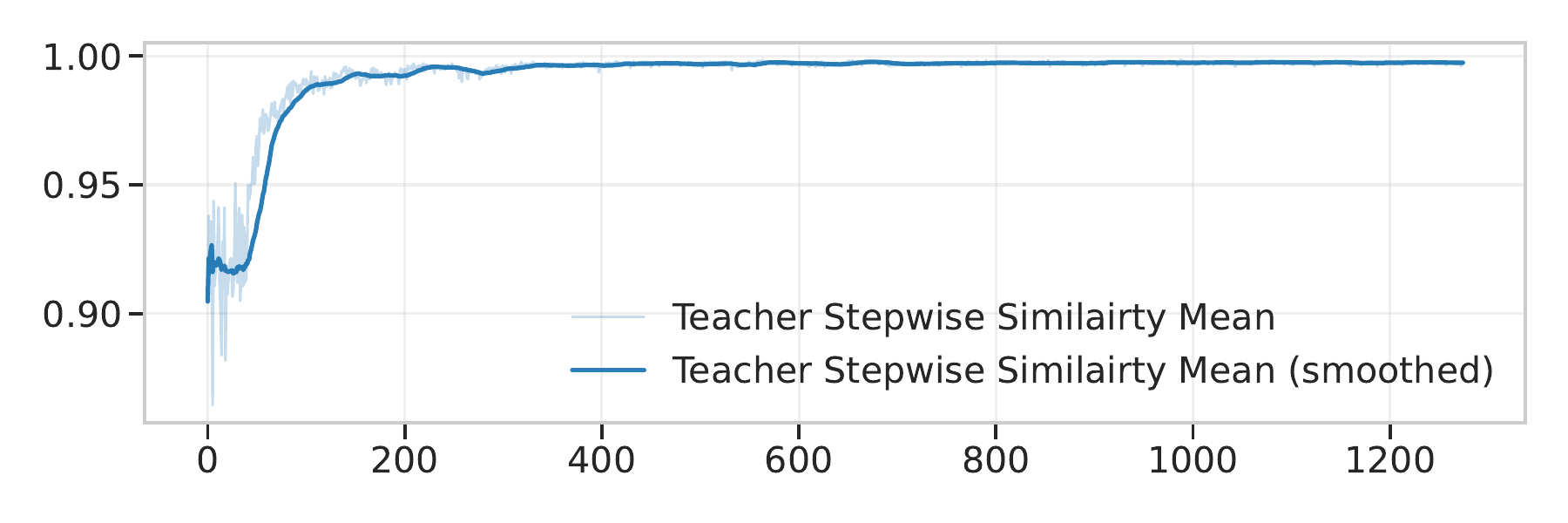}
    \caption{The average similarity of the last hidden states of each explicit reasoning step}
    \label{fig:teacher_sim}
\end{figure}

Our analysis suggests that this high similarity is introduced by LaSER's explicit-view alignment. Although the alignment objective is designed to match different latent representations with different steps of the explicit chain-of-thought trajectory, we find that the teacher model's hidden states at different explicit reasoning steps are already highly similar, as shown in Figure \ref{fig:teacher_sim}. Such a phenomenon, where hidden states tend to occupy a narrow cone and therefore exhibit unusually high cosine similarity, has been shown in prior work, called \textit{anisotropic}~\citep{ethayarajh2019contextual, gao2019representation, wang2020spectrum}. This is especially plausible for autoregressive language models, where each hidden state is optimized for next-token prediction and must summarize the preceding context for generation. Moreover, optimizing retrieval and generation simultaneously under the same representations may be contradictory~\citep{liu2026crem}. Consequently, directly aligning to these hidden states copies the teacher's representational redundancy and the generation noise into the latent reasoning space.

So we first introduce a residual-based retrieval head to extract retrieval-relevant information from the autoregressive hidden states while avoiding the anisotropies. We further adopt bidirectional attention pooling to aggregate the latent states, allowing the model to dynamically assign importance to reasoning representations facing different tasks rather than collapsing them through uniform mean pooling. Finally, we remove all alignment supervision in LaSER. Instead, we design fine-grained self-reconstruction supervision to internalize the alignment via the autoregressive characteristic. 

\subsection{Bi-Attention Aggregates Dynamically}
\label{sec:agg}



\begin{figure}[htbp]
    \centering
    \includegraphics[width=1\columnwidth]{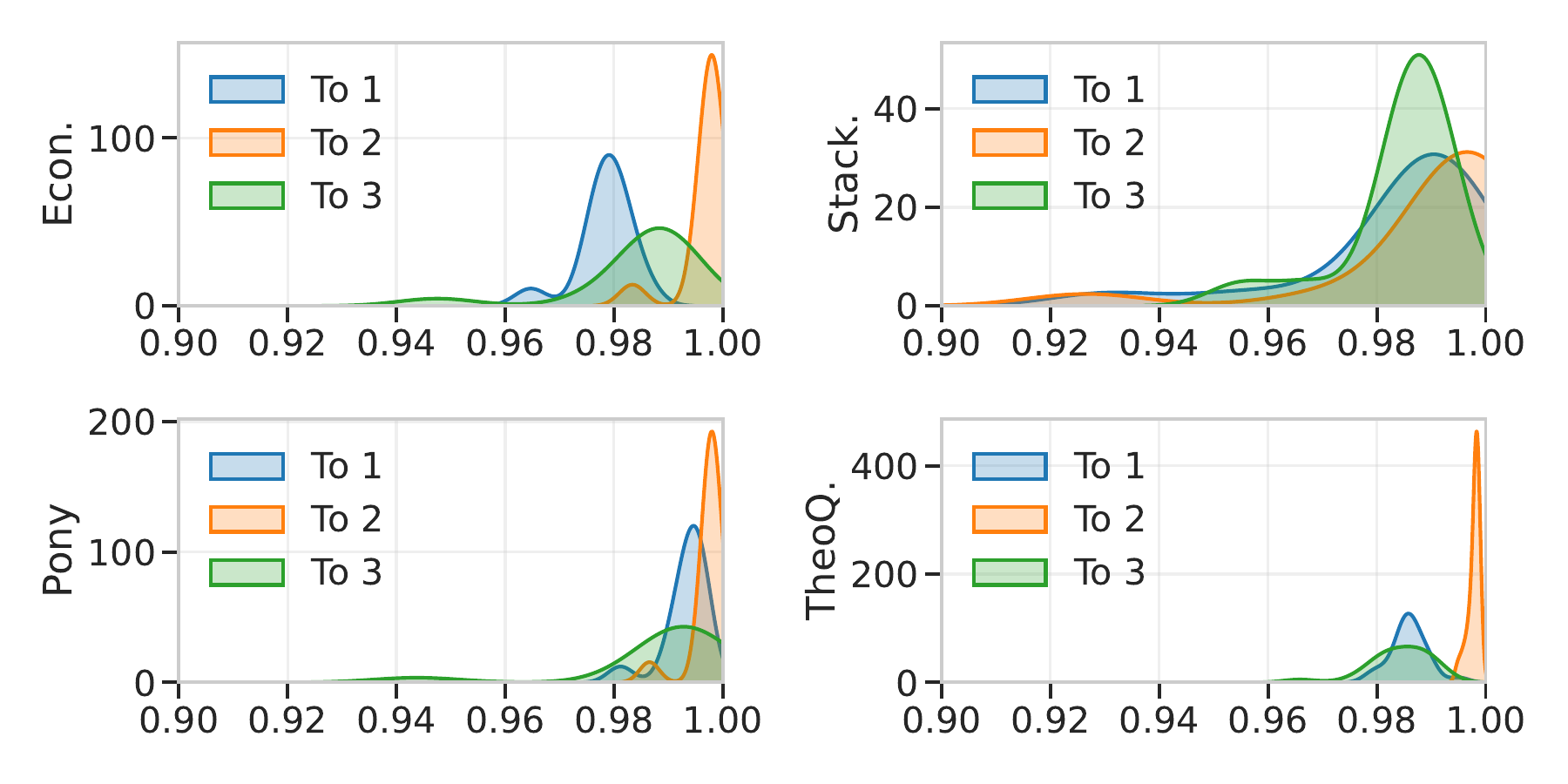}
    \caption{The similarity between the final pooled representation and each latent representation of different tasks in Bright (Full version in the Appendix \ref{sec:app_anaysis}).}
    \label{fig:attn}
\end{figure}

We investigate several aggregation strategies to transfer the reasoning space to the retrieval space, including mean pooling, weighted pooling (fixed for 2:5:10), last pooling, GRU~\citep{cho2014rnn}, causal multi-head attention~\citep{NIPS2017_transformer}, and bidirectional multi-head attention. Empirical results are shown in Table \ref{tab:ablation}: mean pooling introduces redundancy, which leads to sub-optimal performance. Last pooling, weighted pooling, GRU, and causal attention all carry a monotonic prior like GIRCSE that later steps should be more important, which is not always true. Their performance depends on the level of interaction between latent representations. Bidirectional attention is more flexible because it can dynamically assign different weights according to the input distribution. This is consistent with our step-wise retrieval analysis, as Figure \ref{fig:step_perf} shows, where the best-performing reasoning step varies across datasets. And the attention relationship in figure \ref{fig:attn} intuitively illustrates how our approach delivers this capability.

\subsection{Self-Reconstruction Supervises Better}
\label{sec:self_rec}

A common way to impose reconstruction supervision is to attach a lightweight decoder on top of the main encoder representation and require it to reconstruct the original text~\citep{xiao2022retromae}. Following this line, we also compare with a reconstruction baseline that uses a one-layer BERT-style decoder to reconstruct the explicit CoT from the implicit CoT representations. As shown in Table \ref{tab:ablation}, our prediction-based reconstruction performs better. The key distinction is that our method does not treat reconstruction merely as an auxiliary objective imposed by an external decoder. Instead, it builds an internal bridge between abstract implicit reasoning states and concrete explicit reasoning trajectories. 

We further investigate different expansion strategies. One natural alternative is inspired by prefix-tuning~\citep{li2021prefix,li2025coma}. Specifically, we use the $K$ implicit reasoning representations as a prefix, concatenate them with position-initialized slots, and train the slots to fit the explicit CoT. As shown in Table \ref{tab:ablation}, our fine-grained reconstruction provides a more efficient and retrieval-aligned form of supervision.

\subsection{Document-side Reasoning Is Redundant}
\label{sec:ab_doc}

In our framework, reasoning is only applied to the query side. This asymmetry is intentional. Queries are goal-directed and often require reasoning to clarify the underlying information need, whereas documents are static semantic objects without an explicit intent. Our ablation in Table \ref{tab:ablation} shows that applying implicit reasoning to documents, as in LaSER, is a waste when proper supervision is absent. It only increases document encoding cost, which is undesirable because documents are usually encoded at a large scale for indexing.

\section{Conclusion}

In this work, we study how to internalize the reasoning ability of LLM-based retrievers using latent representations. We show that existing implicit-reasoning retrievers still suffer from two mismatches: indirect supervision. To address these issues, we propose \ours, a framework that transfers LLMs into efficient reasoning-intensive retrievers. First, we introduce residual projection and bidirectional attention aggregation to adapt the reasoning states for retrieval. Second, we supervise implicit reasoning with fine-grained next-token-prediction-based reconstruction, shaping latent states with token-level reasoning supervision while contributing to retrieval capability. Experiments on both in-domain and out-of-domain benchmarks show that \ours consistently outperforms standard dense retrievers, ``rewrite-then-retrieve'' pipelines, and comparable latent-reasoning retrievers. Further analyses confirm the importance of properly projecting, aggregating, and supervising latent reasoning states, suggesting a practical direction for efficient reasoning-intensive retrieval.

\section*{Limitations}

\ours has several limitations. First, our training still relies on explicit reasoning trajectories generated during data construction. Their quality, style, and domain coverage can affect the learned latent reasoning space and remain to be researched. Second, \ours uses a fixed number of latent reasoning steps and uniformly maps explicit reasoning tokens to these steps for reconstruction. This design keeps training simple and stable, but real queries may require different reasoning depths. A more adaptive mechanism, such as dynamic latent-step allocation or query-dependent early stopping, may further improve both effectiveness and efficiency.



Finally, latent states are less directly interpretable than textual rationales, so future work should explore better diagnostics for understanding when and how implicit reasoning benefits retrieval.



\bibliography{custom}

\newpage
\appendix

\label{sec:appendix}




\section{Prompt}

The instruction used to expand queries is shown in the following box.

\begin{tcolorbox}[
  enhanced,
  colback=white,
  colframe=black,
  boxrule=0.8pt,
  arc=0pt,
  left=8pt,
  right=8pt,
  top=8pt,
  bottom=8pt
]
\textbf{Instruction:}

Given a question, your mission is to follow the instructions below:

1. Identify the essential problem.

2. Think step by step to reason and describe what information could be relevant and helpful to address the questions in detail.

3. Draft an answer with as many thoughts as you have.

\textbf{The given question:}

\begin{flushleft}
\ttfamily
[Begin of Question]

\textcolor{blue}{\{Query\}}

[End of Question]
\end{flushleft}
\end{tcolorbox}

\newpage
\section{Out-of-Domain Results}
\label{sec:app_results}

The results of FollowIR and BrowseComp-Plus are reported in Table \ref{tab:main2}

\begin{table*}[t]
\centering
\small
\setlength{\tabcolsep}{4.2pt}
\renewcommand{\arraystretch}{1.12}
\caption{Main retrieval performance on follow-up out-of-distribution benchmarks. The best results among models with the same backbone are formatted in bold.}
\resizebox{\textwidth}{!}{%
\begin{tabular}{l c c c c c c c c c c c c}
\toprule
\multirow{2}{*}{\textbf{Models}} & \multirow{2}{*}{\textbf{Size}} &
\multicolumn{2}{c}{\textbf{Core17}} &
\multicolumn{2}{c}{\textbf{News21}} &
\multicolumn{2}{c}{\textbf{Robust04}} &
\multicolumn{2}{c}{\textbf{FollowIR-Avg}} &
\multicolumn{3}{c}{\textbf{BrowseComp-Plus}} \\
\cmidrule(lr){3-4} \cmidrule(lr){5-6} \cmidrule(lr){7-8} \cmidrule(lr){9-10} \cmidrule(lr){11-13}
& &
\textbf{MAP@5} & \textbf{p-MRR} &
\textbf{nDCG@5} & \textbf{p-MRR} &
\textbf{MAP@5} & \textbf{p-MRR} &
\textbf{Score} & \textbf{p-MRR} &
\textbf{R@5} & \textbf{R@100} & \textbf{R@1000} \\
\midrule
\multicolumn{13}{l}{\textit{\textbf{Standard Dense Retrievers}}} \\
BGE-M3 & 0.6B & 1.5 & -1.6 & 21.4 & -1.3 & 7.5 & -8.8 & 10.1 & -3.9 & 4.1 & 21.7 & 47.2 \\
Multilingual-E5-Large-Instruct & 0.6B & 2.4 & 2.7 & 25.0 & 1.2 & 8.5 & -6.0 & 12.0 & -0.7 & 10.1 & 36.8 & 68.6 \\
E5-Mistral-7B-Instruct & 7B & 2.7 & 2.5 & 28.8 & 0.8 & 13.7 & -7.8 & 15.1 & -1.5 & 9.3 & 37.1 & 70.2 \\
Qwen3-Embedding-0.6B & 0.6B & 2.8 & \textbf{8.9} & \textbf{27.3} & \textbf{3.6} & \textbf{11.4} & \textbf{2.7} & \textbf{13.8} & \textbf{5.1} & \textbf{8.0} & \textbf{29.1} & \textbf{64.8} \\
Qwen3-Embedding-8B & 8B & 3.1 & 6.2 & \textbf{25.5} & \textbf{8.8} & 10.1 & \textbf{6.6} & 12.9 & \textbf{7.2} & 7.7 & 31.6 & 61.3 \\
\midrule
\multicolumn{13}{l}{\textit{\textbf{Basic Contrastive Learning}}} \\
Fair Baseline (Qwen3-0.6B) & 0.6B & 2.1 & 3.5 & 13.4 & -0.6 & 6.7 & -3.4 & 7.4 & -0.2 & 3.5 & 21.3 & 46.9 \\
Fair Baseline (Qwen3-8B) & 8B & 2.8 & 4.4 & 18.9 & 2.0 & 11.2 & -1.3 & 11.0 & 1.7 & 11.3 & 37.4 & 63.2 \\
Fair Baseline (LLaMA3.1-8B) & 8B & 2.5 & \textbf{2.8} & 18.9 & 0.1 & 8.1 & -2.6 & 9.8 & 0.1 & 6.1 & 25.4 & 50.8 \\
\midrule
\multicolumn{13}{l}{\textit{\textbf{Explicit Reasoning}}} \\
Search-R3 (Qwen2.5-1.5B) & 1.5B & 3.2 & 7.1 & 26.2 & -0.1 & 10.1 & 2.7 & 13.2 & 3.2 & 0.0 & 0.3 & 1.1 \\
\midrule
\multicolumn{13}{l}{\textit{\textbf{Latent Reasoning}}} \\
GIRCSE (Qwen3-0.6B) & 0.6B & \textbf{3.0} & 3.9 & 12.8 & -0.9 & 5.7 & -7.7 & 7.2 & -1.6 & 6.9 & 25.2 & 52.8 \\
GIRCSE (Qwen3-8B) & 8B & 3.0 & 4.2 & 22.6 & 0.7 & 8.5 & 1.0 & 11.4 & 2.0 & 13.0 & \textbf{40.8} & \textbf{68.1} \\
GIRCSE (LLaMA3.1-8B) & 8B & 2.9 & 1.0 & 19.3 & -1.3 & \textbf{12.2} & \textbf{-1.2} & 11.5 & -0.5 & 6.7 & 23.5 & 51.0 \\
LaSER(Qwen3-0.6B) & 0.6B & 2.0 & 3.4 & 25.0 & 1.0 & 7.4 & -4.5 & 11.5 & 0.0 & 6.8 & 26.8 & 54.9 \\
LaSER(Qwen3-8B) & 8B & 4.1 & 5.8 & 21.8 & 0.2 & 11.4 & 1.3 & 12.5 & 2.4 & 11.7 & 38.4 & 66.9 \\
LaSER(LLaMA3.1-8B) & 8B & 3.6 & 2.7 & 22.0 & \textbf{1.6} & 11.1 & -1.9 & 12.2 & \textbf{0.8} & \textbf{6.8} & 25.7 & \textbf{52.9} \\
\rowcolor{blue!10}
\textbf{\ours} (Qwen3-0.6B) & 0.6B & 2.0 & 6.7 & 23.1 & 0.2 & 8.9 & -2.2 & 11.3 & 1.6 & 6.9 & 27.4 & 56.1 \\
\rowcolor{blue!10}
\textbf{\ours} (Qwen3-8B) & 8B & \textbf{4.3} & \textbf{6.4} & 24.6 & 0.4 & \textbf{12.7} & 0.8 & \textbf{13.9} & 2.5 & \textbf{13.2} & 40.6 & 66.8 \\
\rowcolor{blue!10}
\textbf{\ours} (LLaMA3.1-8B) & 8B & \textbf{4.1} & 2.5 & \textbf{23.0} & 0.6 & 10.8 & -1.7 & \textbf{12.6} & 0.5 & \textbf{6.8} & \textbf{26.9} & 52.7 \\
\bottomrule
\end{tabular}%
}
\label{tab:main2}
\end{table*}

\section{Analysis}
\label{sec:app_anaysis}

\begin{figure}[htbp]
    \centering
    \includegraphics[width=1\columnwidth]{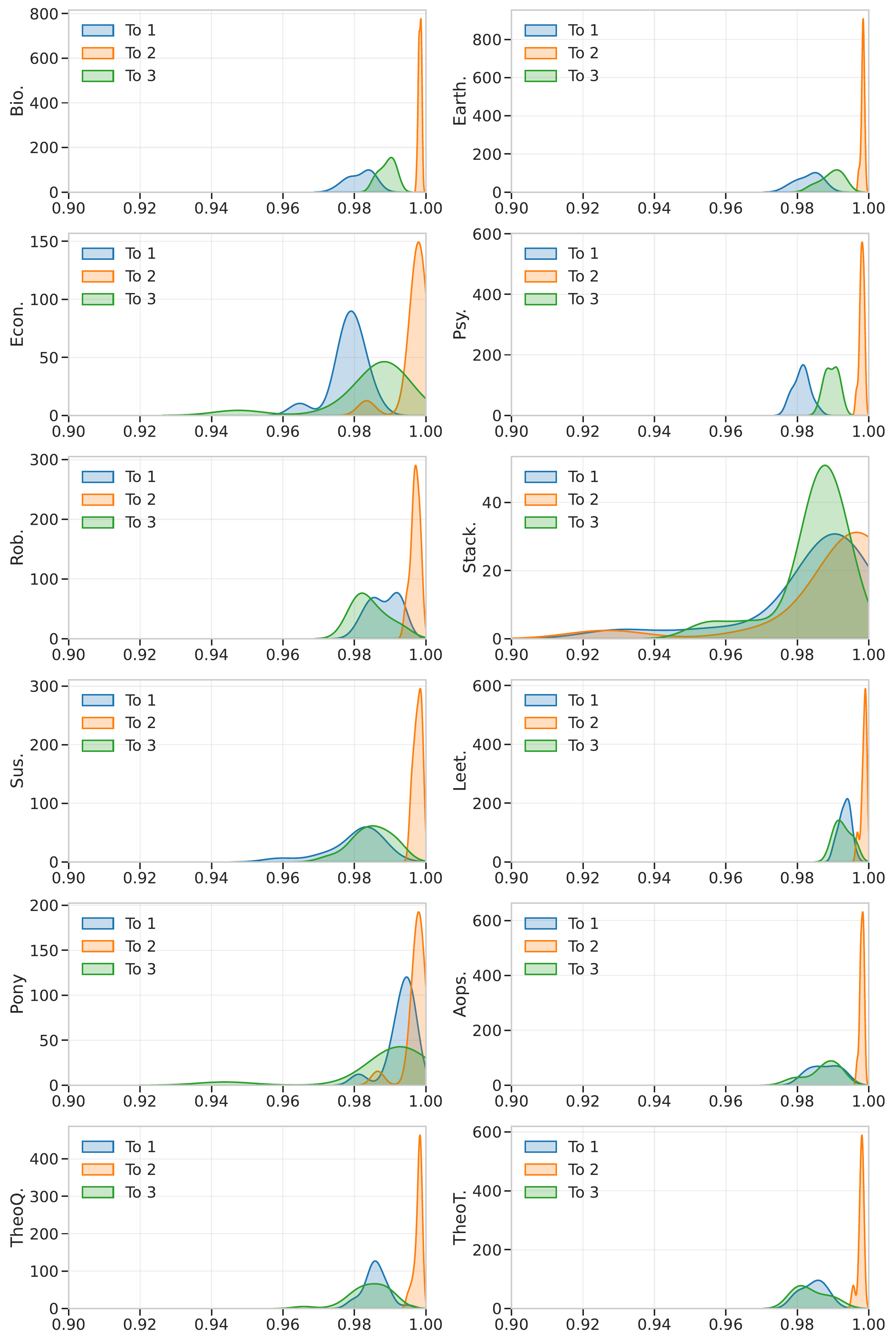}
    \caption{The similarity between the final pooled representation and each latent representation of different tasks in Bright}
    \label{fig:attn_full}
\end{figure}

The full version of Figure \ref{fig:attn} in Section \ref{sec:agg} in shown in Figure \ref{fig:attn_full}

\end{document}